\documentclass[
    ,final            
  ]
  {aipproc}

\layoutstyle{8x11single}
\graphicspath{{./images/}}
\usepackage{multirow}

\begin{document}

\title{Measurement of the transverse Single Spin Asymmetry of Forward Eta Mesons in $p^{\uparrow}+p$ Collisions in PHENIX}


\author{David Kleinjan for the PHENIX collaboration}{
  address={University of California, Riverside}
}

\begin{abstract}
The measurement of inclusive meson transverse single spin asymmetries ($A_N$) from transversely polarized $p+p$ collisions provides insight into the structure of the nucleon. $A_N$ of eta mesons at forward rapidities are of particular interest, as asymmetries of other mesons observed both by PHENIX and other experiments show significant deviation from zero in the forward region.  Several mechanisms have been proposed that attempt to explain these non-zero asymmetries, and a comparison of different probes may further constrain these models. Therefore, measurements of $A_N$ with inclusive eta mesons at forward rapidities are an important tool toward understanding the underlying mechanism. Using the PHENIX detector at the Relativistic Heavy Ion Collider (RHIC), we study $p^{\uparrow}+p$ collisions. In 2008, the PHENIX experiment collected 5.2 pb$^{-1}$ integrated luminosity in $p^\uparrow+p$ collisions at $\sqrt{s}$ = 200 GeV.  The asymmetry analysis of eta mesons at forward rapidity will be discussed.
\end{abstract}

\maketitle

\section{Introduction}
\begin{figure}[h!]
\includegraphics[width=25pc]{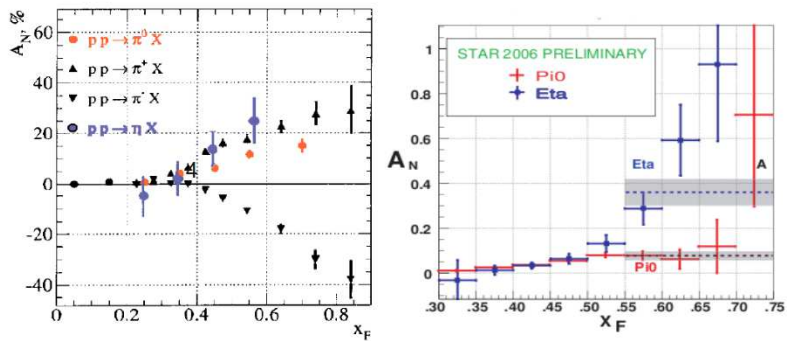}\hspace{1pc}
\caption{\label{fig:etaan}Two previous measurements of eta meson transverse single spin asymmetry.  The left figure shows results from FNAL-E704 at $\sqrt{s}$ = 19.4 GeV \cite{bibeta1}.  The right figure shows preliminary results from the STAR experiment at $\sqrt{s}$ = 200 GeV \cite{bibeta2}.}
\end{figure}

Inclusive mesons from transversely polarized $p^\uparrow+p$ collisions have an analyzing power, $A_{N}$, which provides insight into the composite structure of the proton.  The analyzing power, or transverse single spin asymmetry, is quantified  by taking the ratio of the difference and sum of independent polarized processes:
\begin{equation}
\displaystyle A_{N} = \frac{f}{P} \frac{(\sigma^{\uparrow} - \sigma^{\downarrow})}{(\sigma^{\uparrow} + \sigma^{\downarrow})}
\label{eq:an_equ}
\end{equation}
where $P$ is the average beam polarization, $f$ is a geometric scale factor that corrects for the azimuthal acceptance of the detector, and $\sigma^\uparrow$($\sigma^\downarrow$) is the meson production cross section from $p^\uparrow$($p^\downarrow$) + $p$ beam crossings. In prior measurements, non-zero transverse single spin asymmetries have been measured for charged \cite{bibpion0,bibpion1,bibpion2,bibpion3,bibpion4,bibpion5} and neutral \cite{bibpi00, bibpi01, bibpi02} pions at large Feynman-$x$ ($x_{F}~=~2p_{L}/\sqrt{s}$, where $p_{L}$ is the momentum of the meson along the beam direction) at various collision energies. The transverse single spin asymmetry of eta mesons have also been measured by the FNAL-E704 \cite{bibeta1} and STAR \cite{bibeta2} experiments, shown in Fig. \ref{fig:etaan}.  FNAL-E704 measured similar asymmetries for neutral pions and eta mesons at $\sqrt{s}$ = 19 GeV; STAR measured a larger asymmetry for eta mesons compared to neutral pions at $x_F~>~$0.55.

Since collinear pQCD at leading twist predicts a small spin asymmetry due to the hard partonic scattering, initial and final states of the interacting partons and fragmenting hadrons have to be considered.  Three main explanations have been developed to explain these transverse single spin asymmetries:  the Sivers effect \cite{sivers}, the Collins effect \cite{collins}, higher twist effects \cite{twist}, or some combination of the three.  

The goal of this analysis is the extraction of the inclusive eta meson transverse single spin asymmetry at forward rapidity at PHENIX \cite{phenix}.  In 2008, the PHENIX experiment collected 5.2 pb$^{-1}$ integrated luminosity in $p^\uparrow+p$ collisions at $\sqrt{s}$ = 200 GeV, with an average beam polarization of 45$\%$.

\section{Measurement}
For the analysis of the forward eta meson transverse single spin asymmetry in PHENIX, the Muon Piston Calorimeter (MPC) is being used\cite{mpc}.  The MPC consists of two forward electromagnetic calorimeters, referred to as the south (north) MPC, placed $\pm$220 cm from the nominal interaction point at a pseudorapidity of 3.1 $<~\eta~<$ 3.9.  Each MPC comprises 196 (220) $PbWO_{4}$ crystal towers, with dimensions of $2.2 \times 2.2 \times 18~$cm$^{3}$.  The MPC can identify neutral pions and eta mesons by reconstruction of their decay photons.  All pairs of photon candidates are used to form an invariant mass.  Real photon pairs originating from eta meson decays will form a pair-mass close to the mass of the eta meson.   PHENIX operates within the RHIC accelerator, which provides two counter-circulating proton beams at various energies up to $\sqrt{s}$ = 500 GeV.  A feature of RHIC is its capability to provide polarized proton beams independently in both directions, allowing for two independent transverse single spin asymmetry measurements by integrating over one beam polarization at a time.

\begin{figure}[h]
\includegraphics[width=13pc]{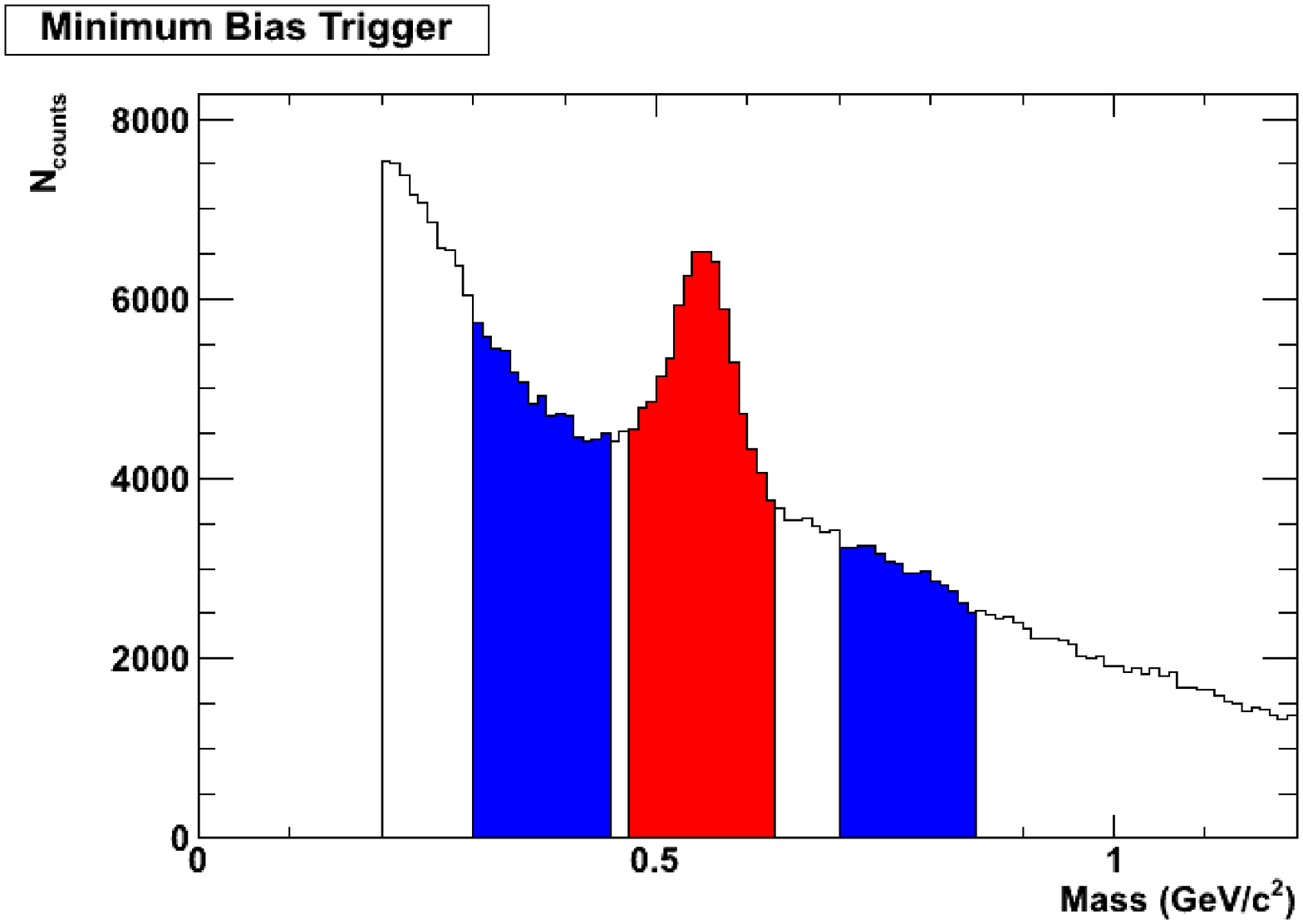}\hspace{2pc}%
\includegraphics[width=13pc]{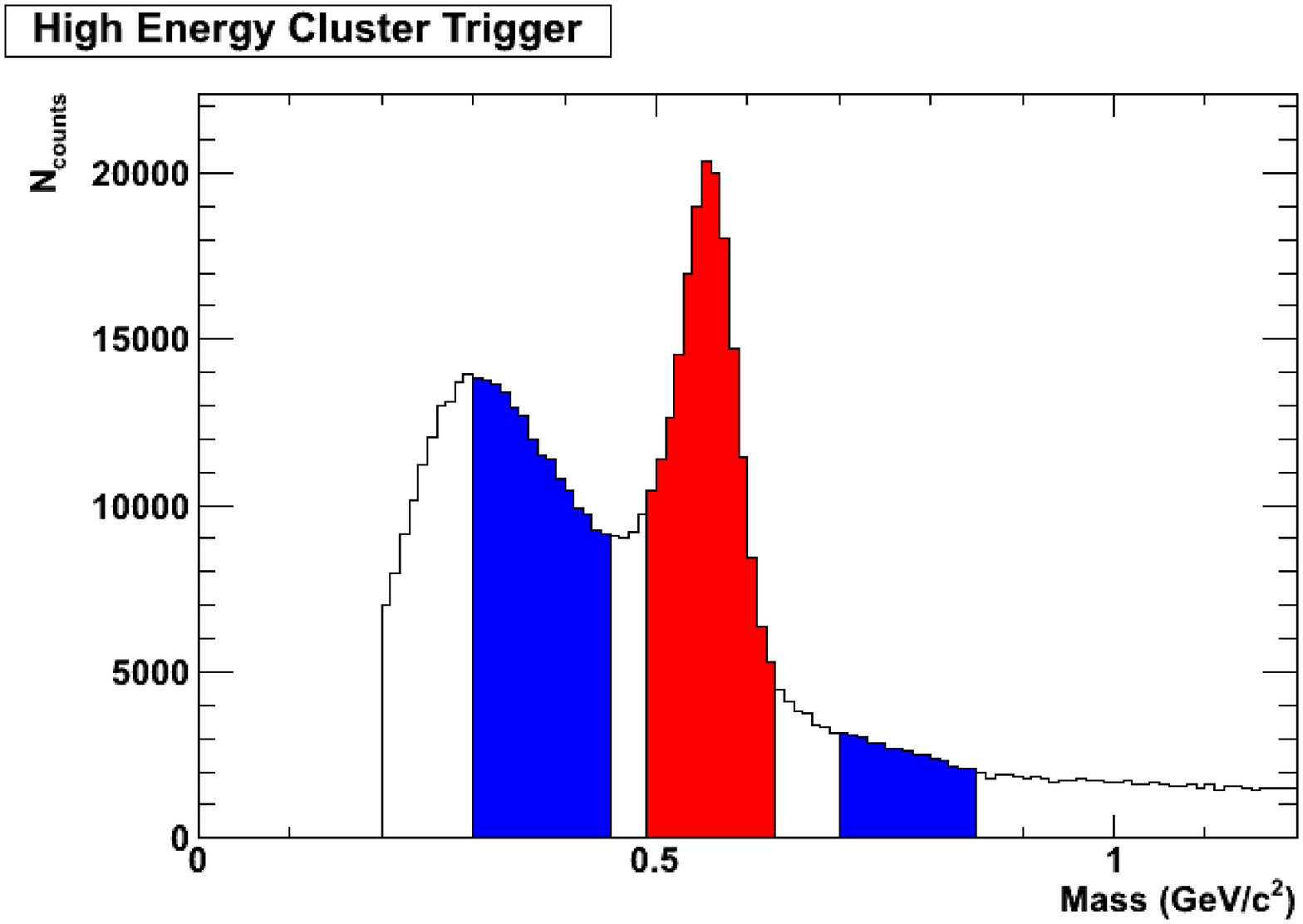}\hspace{2pc}%
\caption{\label{fig:sideband} The invariant mass distributions from MB (left) and HEC (right) trigger data. The asymmetry, $A_N$, is calculated in three different regions: the eta meson peak region (red), as well as the sideband regions (blue).}
\end{figure}

In order to record the data, two triggers are used: a minimum bias (MB) trigger which uses the Beam-Beam Counters only, and a high energy cluster (HEC) trigger in which a $4 \times 4$ tower energy sum (threshold of $E_{4 \times 4}$ $\sim$ 20 GeV) in the MPC is used to initiate readout of the detector.  The MB trigger allows for measurement of the eta meson yield in the region of 1 $<~p_{T}~<$ 2 GeV/$c$ (0.2 $<~x_F~<$ 0.4), where $p_T$ is the transverse momentum of the eta meson normal to the beam direction.  The HEC triggered dataset allows reconstruction of eta mesons in the region 2 $<~p_{T}~<$ 4.5 GeV/$c$ (0.3 $<~x_F~<$ 0.7).  Figure~\ref{fig:sideband} shows the invariant mass distribution for the different trigger types.

The invariant mass distributions in Fig. \ref{fig:sideband} show a clear eta meson peak in the region of the eta meson mass of 0.548 GeV/$c^2$ \cite{pdg_ref}.  However, there remains a large combinatorial background under the eta meson peak.  This background is corrected for in the final $A_N$ calculation, using the formula:
\begin{equation}
\displaystyle A_{N} = \frac{A^{inc}_{N} - rA^{bg}_{N}}{1-r}, \displaystyle \sigma_{A_{N}} = \frac{\sqrt{\sigma^{2}_{A^{inc}_{N}} + r^{2} \sigma^{2}_{A^{bg}_{N}}}}{1-r}
\label{eq:ancorr}
\end{equation}
where $r$ is the fraction of background in the eta meson peak region, $A^{inc}_{N}$ is the asymmetry in the eta meson mass region, and $A^{bg}_{N}$ is the asymmetry of the background, which is the weighted mean of the asymmetries in low and high invariant mass sideband regions relative to the eta meson mass peak, $A_N^{low}$ and $A_N^{high}$.  Figure \ref{fig:sideband} shows these different mass regions where $A_{N}^{inc}$, $A_N^{low}$ and $A_N^{high}$ are calculated.

\begin{figure}
\includegraphics[width=0.45\linewidth]{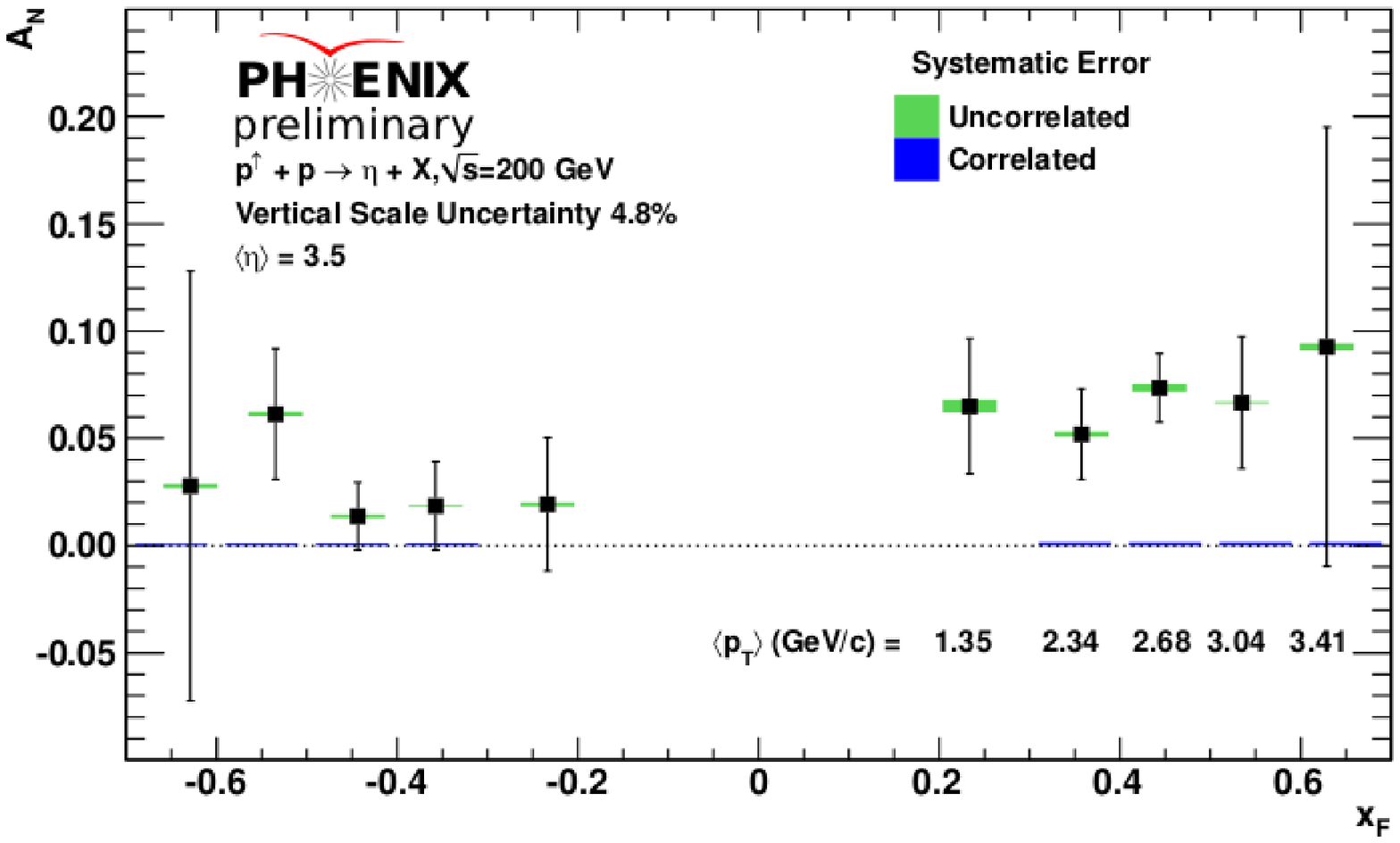}
\includegraphics[width=0.45\linewidth]{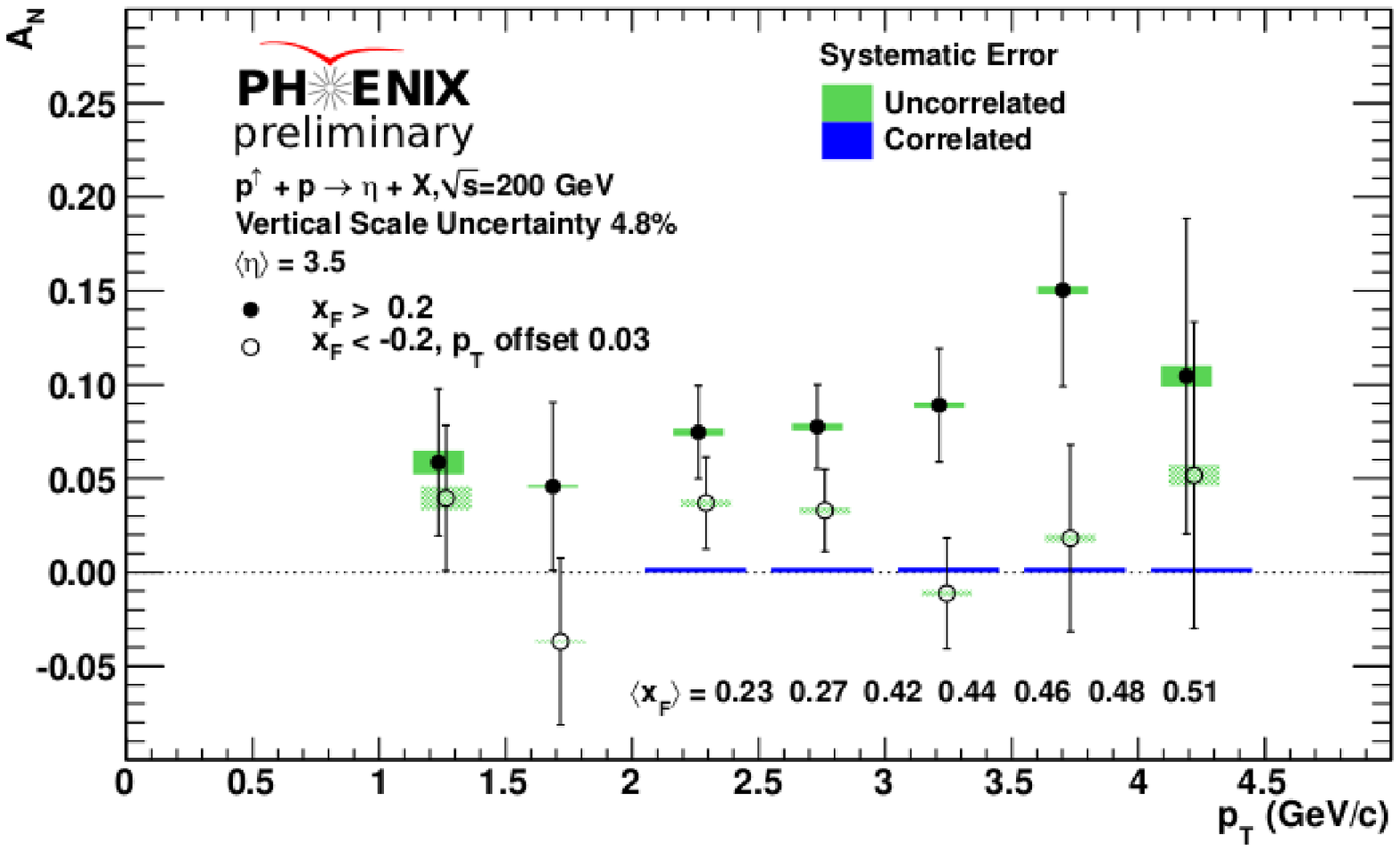} \\
\caption{\label{fig:final}  The asymmetries.  The left (right) panel show the $x_F$ ($p_T$) dependence of the asymmetry.}
\end{figure}

\begin{figure}
\includegraphics[width=0.45\linewidth]{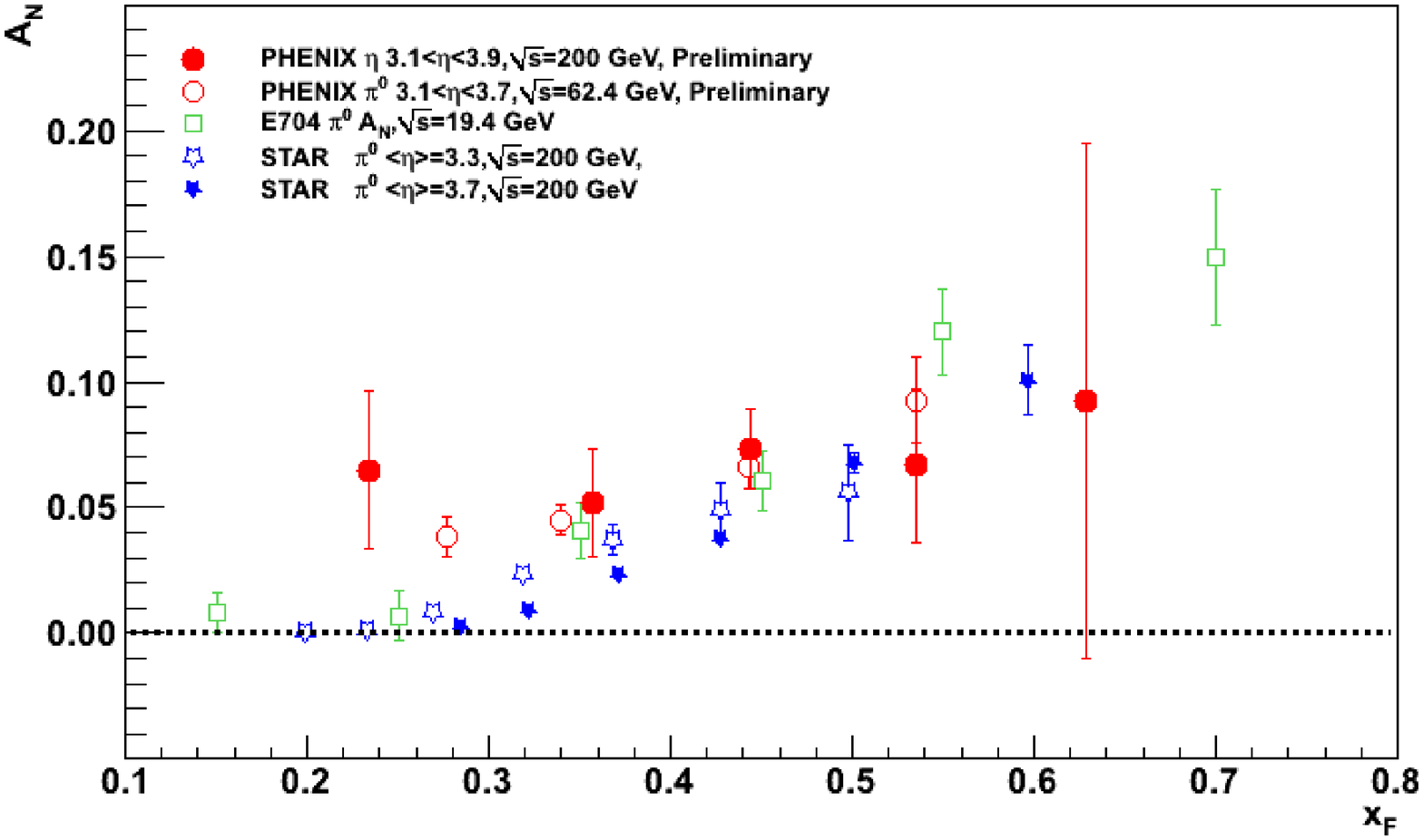}
\caption{\label{fig:topi0s}   Comparison of PHENIX $\eta$ meson asymmetry to other neutral pion asymmetries in the forward region \cite{bibpi00, bibpi01, bibpi02}.}
\end{figure}

The measured asymmetries, $A_N$, are shown in Fig. \ref{fig:final}.  The left panel shows the $x_F$ dependence of the asymmetries.  The weighted mean of the negative (positive) $x_F$ points are 2.2$\pm$1.1$\%$ (6.6$\pm$1.1$\%$), which indicates a negative $x_F$ asymmetry consistent with zero, and a clear non-zero asymmetry at positive $x_F$.  The right panel shows the $p_T$ dependence of the asymmetries.  The open (closed) points are for negative (positive) $x_F$ values with a weighted mean of 2.4$\pm$1.3$\%$ (8.5$\pm$1.4$\%$). There is a small point-to-point uncorrelated systematic error based on the determination of $r$ in eq. \ref{eq:ancorr}, and a small correlated systematic error based on variations of the polarization of the $p+p$ collisions.  Figure \ref{fig:topi0s} compares the positive $x_F$ dependence of $A_N^{\eta}$ to previous measurements of $A_N^{\pi^0}$, which demonstrates that this new measurement of the transverse single spin asymmmetry of eta mesons is consistent with previous measurements of neutral pions.

\bibliographystyle{aipproc}   

\end{document}